\documentclass[preprint, double-spaced]{revtex4}
 \usepackage{wrapfig}
\usepackage{graphicx}
\usepackage{amsmath,bbm}
\usepackage{amssymb,bm}

\begin{document}

\title{A Hybrid Model for QCD Deconfining Phase Boundary }
\author{P.~K.~Srivastava}
\author{C.~P.~Singh}

\affiliation{Department of Physics, Banaras Hindu University, 
Varanasi 221005, INDIA}

\begin{abstract}

Intensive search for a proper and realistic equations of state (EOS) is still continued for studying the phase diagram existing between quark gluon plasma (QGP) and hadron gas (HG) phases. Lattice calculations provide such EOS for the strongly interacting matter at finite temperature ($T$) and vanishing baryon chemical potential ($\mu_{B}$). These calculations are of limited use at finite $\mu_{B}$ due to the appearance of notorious sign problem. In the recent past, we had constructed a hybrid model description for the QGP as well as HG phases where we make use of a new excluded-volume model for HG and a thermodynamically-consistent quasiparticle model for the QGP phase and used them further to get QCD phase boundary and a critical point. Since then many lattice calculations have appeared showing various thermal and transport properties of QCD matter at finite $T$ and $\mu_{B}=0$. We test our hybrid model by reproducing the entire data for strongly interacting matter and predict our results at finite $\mu_{B}$ so that they can be tested in future. Finally we demonstrate the utility of the model in fixing the precise location, the order of the phase transition and the nature of CP existing on the QCD phase diagram. We thus emphasize the suitability of the hybrid model as formulated here in providing a realistic EOS for the strongly interacting matter.
\\

 PACS numbers: 12.38.Mh, 12.38.Gc, 25.75.Nq, 24.10.Pa

\end{abstract}
\maketitle 
\section{Introduction}
\noindent
Relativistic heavy ion collider (RHIC) at Brookhaven National Laboratory (BNL), large hadron collider (LHC) at CERN and future compressed baryonic matter (CBM) experiments at GSI are providing a great opportunity to study the properties of strongly interacting matter in the laboratory at extreme temperatures and baryon densities. The outcomes of these experiments will be of an enormous use in obtaining a proper equation of state (EOS) for high and low temperature QCD phases [1-3]. On the theoretical side, lattice calculations using QCD thermodynamics provide a valid EOS for strongly interacting QCD matter but the method still lacks reliability for a matter possessing a finite density of baryons. Therefore, precise mapping of the entire QCD phase boundary and the location of a hypothesized QCD critical point (CP) still pose a challenging problem before the experimental and theoretical heavy-ion physicists today [4-6]. Lattice people use certain approximations in obtaining thermodynamical quantities at small baryon chemical potential ($\mu_{B}$) and indicate the existence of a cross-over chiral transition at $\mu_{B}= 0$ and $T \approx 170$ MeV which ends at a CP for $\mu_{c}/3T_{c}\le 1.0$ [7-8]. However, some recent lattice results showed doubts over the existence of a CP on the QCD phase boundary [9]. We still do not precisely know whether the conjectured phase boundary is an outcome of deconfinement and/or chiral phase transition and whether CP appears only on the chiral phase boundary [10]. In such circumstances, it seems worthwhile to intensify the search for a realistic EOS which can suitably describe both the QCD phases. In this paper, our aim is to obtain such a description using phenomenological models for the QGP as well as HG phases separately and combine them suitably in a hybrid model formulation. We then proceed to test whether the model results exactly reproduce the most recent lattice data which have arrived for vanishing $\mu_{B}$ and at finite temperature and thereafter the precise location and the nature of the phase transition at CP are investigated in this model.

 Recently several attempts were made to explain the lattice data at low temperatures by using ideal hadron gas (IHG) description where interactions among hadrons are altogether neglected [11-15]. However, Andronic and collaborators indicated in a most recent paper that the features of the low energy lattice data arise if a hard-core repulsion among the various constituents of the hadron gas is incorporated as excluded volume effect [16]. Simultaneously, several papers addressed the high temperature lattice data by using a suitable quasiparticle description for QGP in which the constituents of QGP acquire a $T$ and/or $\mu_{B}$- dependent mass [17-19]. These results suggest us to use a hybrid model type description in which low temperature phase can be described by a thermodynamically - consistent excluded volume HG model and high temperature QGP phase is suitably described by a thermodynamically consistent quasiparticle model. Recently we have used this hybrid model in constructing a first order deconfining phase boundary between HG and QGP by employing Gibbs' equilibria conditions and found that the boundary indeed terminates at a CP beyond which a crossover region exists [20, 21]. The main difference between our model and the one used by Andronic et al [16] is that we provide a hard-core size to baryons only and mesons in our model can overlap and penetrate into each other whereas Andronic et al have given the same hard core size to all the hadrons existing in HG. Recently we have demonstrated that our model successfully describes the multiplicity distributions, multiplicity ratios, shear viscosity to entropy density ($\eta/s$) ratio, square of speed of sound ($c_{s}^{2}$) etc and our results compare well with the experimental data [22]. There are various thermodynamically consistent versions of quasiparticle model existing in the literature [17-19]. However, our quasiparticle description involves only two parameters i.e., $\Lambda$ and $T_{0}$ appearing in the effective coupling constant [21] in comparison to other approaches which have three or four parameters.
 
 We must emphasize the new results obtained by us in this paper in order to demonstrate its importance. In particular, we first calculate various thermodynamical quantities e.g., three times the normalized pressure density $3p/T^{4}$, normalized energy density $\epsilon/T^{4}$, normalized entropy density $s/T^{3}$, trace anomaly factor ($\epsilon-3p)/T^{4}$ revealing a measure of the interaction present in the QCD matter and the normalized baryonic succeptibility ($\chi^{B}_{2}/T^{2}$) etc for which the lattice data have recently become available at $\mu_{B}=0$ [11, 13, 17, 23-24]. We show their variations with temperature and compare our curves with the lattice data for the entire QCD matter. The comparisons thus demonstrate the validity of our hybrid model in providing a realistic description of both the phases of QCD matter. We also extend our studies of the above quantities for finite $\mu_{B}$ so that our results can be put to test when the lattice calculations in future become feasible. Similarly we also show results for the ratio of pressure to energy ($p/\epsilon$), square of the speed of sound ($c_{s}^{2}$) and other transport ratios like shear viscosity to entropy density ($\eta/s$) for the gluon plasma only and compare them with the available lattice results at $\mu_{B}=0$. Moreover, in this paper, we get the location of CP and determine the order of the phase transition by studying the difference in entropy density i.e., $\Delta s/T^{3}$ as well as difference in sound speed $\Delta c_{s}^{2}$ between HG and QGP phases exactly at the phase boundary. We notice that these quantities vanish at the critical point and hence indicate a clear change in the order of the phase transition at CP. Recent studies of Csernai at al [25], Sasaki and Redlich [26] and Lacey et. al. [27] have revealed that $\eta/s$ ratio involves a cusp like feature in the graph near the critical point when we plot its variation with temperature $T$. Our results in hybrid model supports these findings and thus helps in precisely locating the critical point.

 The rest of the paper is organized as follows : In Sec. II, we give a brief outline for the EOS in QGP phase and a detailed prescription we have used to calculate various thermodynamical as well as transport properties. In Sec. III, we obtain the EOS for HG in the new excluded-volume model and derive the relations to calculate various thermodynamical and transport properties from it. In Sec. IV, we present a detailed comparison of our results with those from recent lattice calculations and also predict the variations of these quantities at finite $\mu_{B}$. Finally, Sec. V fixes the detailed comparisons of results regarding the order of the phase transition at CP on the deconfining phase boundary between HG and QGP and final conclusions are then presented.
\section{EOS for QGP}
The EOS for QGP in a quasiparticle framework as used in this paper has been described in detail in Ref. [28, 29] and the calculations regarding thermodynamical quantities like pressure, energy density, particle density etc. can be found in our earlier work [20]. Here we add how our prescription can be used to obtain the transport properties e.g., shear viscosity, speed of sound etc. In this model, we start with the definition of average energy density and average number density of particles and derive all other thermodynamical quantities from them in a consistent manner. The expressions for energy density, number density and pressure are [20]:

\begin{equation}
\epsilon=\frac{T^4}{\pi^2}\sum_{l=1}^{\infty}\frac{1}{l^4}\left[\frac{d_g}{2}\epsilon_{g}(x_{g}l)+(-1)^{l-1}d_{q}cosh(\mu_{q}/T)\epsilon(x_{q}l)+(-1)^{l-1}\frac{d_{s}}{2}\epsilon_{s}(x_{s}l)\right],
\end{equation}

\begin{equation}
n_{q}=\frac{d_{q}T^{3}}{\pi^2}\sum_{l=1}^{\infty}(-1)^{l-1}\frac{1}{l^3}sinh(\mu_{q}/T)I_{i}(x_{i}l),
\end {equation}
and,
\begin{equation}
\it{p}(T,\mu_{q})=\it{p}(T,0)+\int_{0}^{\mu_{q}}n_{q}d\mu_{q}.
\end{equation}
In Eq. (3), $\it{p}(T,0)$ is defined as follows [20]:
\begin{equation}
\frac{\it{p}(T,\mu_{q}=0)}{T}=\frac{\it{p}_0}{T_0}+\int_{T_0}^{T}dT \frac{\epsilon(T,\mu_{q}=0)}{T^2}.
\end{equation}
 In Eq. (1), $\epsilon_{i}(x_{i}l)=(x_{i}l)^{3}K_{1}(x_{i}l)+3 (x_{i}l)^{2}K_{2}(x_{i}l)$, where $K_1$ and $K_2$ are the modified Bessel functions with $x_{i}=\frac{m_i}{T}$ and index i runs for gluons, up-down quarks q, and strange quark s. Similarly in Eq. (2), $I_{i}(x_{i}l)=(x_{i}l)^2 K_{2}(x_{i}l)$. $d_i$ are the degeneracies associated with the internal degrees of freedom and $p_{0}$ is the pressure at $T=T_{0}$. Entropy density for QGP can be obtained using the expression:
\begin{equation}
s=\frac{\epsilon + p-\sum_{i}\mu_{i}n_{i}}{T},
\end{equation}
where index i runs for different flavours of quarks. The baryonic succeptibility can be obtained from the following relation [30]:
\begin{equation}
\chi_{2}^{B}=\frac{1}{3}\chi_{q}=\frac{\partial n_{q}}{\partial \mu_{q}}.
\end{equation}
 Let us now calculate the transport properties. Our calculation for shear viscosity is based on the prescription used by Sasaki and Redlich [32] who have calculated shear as well as bulk viscosities for QGP in the quasiparticle model. By definition, shear ($\eta$) and the bulk viscosities ($\zeta$) are defined as coefficients of the space-space component of the deviations of the energy momentum tensor from equilibrium if a small perturbation causes the system to slightly deviate from its equilibrium value. Near equilibrium, one gets the total energy-momentum tensor as follows [31, 32]:

\begin{equation}
T_{\mu\nu}=T_{\mu\nu}^{0}+\Delta T_{\mu\nu},
\end{equation}
here $T_{\mu\nu}^{0}= \int d\frac{d^3k}{(2\pi)^3} {}\frac{k^\mu k^\nu}{E}\left[ f_{0}+ \bar{f_{0}}\right]\,,$ is the energy momentum tensor of the system in equilibrium with $f_{0}$ ($\bar{f}_{0}$) as the equilibrium distribution functions for particles (anti-partciles) and $d$ is the degeneracy factor of particles/anti-particles. Further, $\Delta T_{\mu\nu}$ is the deviation in energy momentum tensor caused by a shift in the equilibrium. We can expand $\Delta T_{\mu\nu}$ in time-time, space-time and space-space part as follows [32]:
\begin{equation}
\Delta T_{\mu\nu}=\Delta T_{00}+\Delta T_{0i}+\Delta T_{i0}+\Delta T_{ij}.
\end{equation}
Now the space-space components of $\Delta T_{\mu\nu}$ can be written as the sum of traceless $W_{ij}$ and the scalar part [31, 32]:
\begin{equation}
\Delta T_{ij}=-\zeta\delta_{ij}\partial_{m}u^{n}-\eta W_{ij},
\end{equation}
which involve the transport coefficients eg, the bulk ($\zeta$) and shear($\eta$) viscosities, respectively. In Eq. (9), $W_{mn}=\left( \partial_m u^n{}+ \partial_n u^m - \frac{2}{3}\delta_{mn}\partial_i u^i \right)$ where $u$ represents the flow velocity. Using relaxation time approximation, the shear viscosity in a medium composed of one type of particles/antiparticles can be obtained from the following expression [31] :
\begin{eqnarray}
&&
\eta =  \frac{1}{15T}\int\frac{d^3k}{(2\pi)^3} 
\frac{k^4}{E^2}~d~\tau
\left[ f_0(1\pm f_0) 
{}+ \bar{f}_0(1\pm\bar{f}_0)  \right]\,, 
\end{eqnarray}
where $k$ is the momentum and $E=\sqrt{{ k}^2+M^2}$ is the energy with $M$ being the thermal mass for quark or gluon, $\pm$ for fermion and boson (i.e., quark and gluon), respectively and $f_{0}$ ($\bar{f}_{0}$) stands for equilibrium distribution function for particle and anti-particle, respectively [31]:
\begin{eqnarray}
f_{0}(\bar{f}_{0}) = (e^{(E \mp \mu)/T}\pm1)^{-1}\,,
\end{eqnarray}
where we use $\mp \mu$ in $f_{0}$ and $\bar{f}_{0}$, respectively. In Eq. (11), we use $\pm 1$ for quark and gluon, respectively. In Eq. (10), $\tau$ is the collision time. To calculate the collision time for quarks, antiquarks and gluons in QGP we use the following expressions [32,33]:
\begin{equation}
\tau_{q(\bar{q})}=\frac{1}{15\alpha_{s}^{2} T{ } log \left(\frac{1}{\alpha_{s}}\right)\left(1+0.06 N_{F}\right)},
\end{equation}
\begin{equation}
\tau_{g}=\frac{1}{3.4\alpha_{s}^{2} T { } log \left(\frac{1}{\alpha_{s}}\right)\left(1+0.12 (2 N_{F}+1)\right)}.
\end{equation}
Here $\alpha_{s}=g^{2}/4\pi$, is QCD running coupling constant where $g^{2}$ has been taken from Ref. [20] and $N_{F}$ is the number of effective flavour degrees of freedom. We get the final expression for shear viscosity by adding the contributions of all types of particles in Eq. (10). Similarly the speed of sound in QGP is :
\begin{equation}
c_{s}^{2}=\left(\frac{\partial p}{\partial \epsilon}\right)_{s/n}=\left(\frac{\partial p/\partial T}{\partial \epsilon/\partial T}\right)_{s/n},
\end{equation}
where $p$ is the pressure and $\epsilon$ is the energy density of the QGP as given in Eq. (3) and (1), respectively.\\
\section{EOS for HG}
 Recently we proposed a new thermodynamically consistent, excluded-volume model for the hot and dense HG [20-22, 34]. Our approach incorporates the following new features. Besides thermodynamical consistency, our model uses full quantum statistics so that our model will remain valid even in the case of large $\mu_{B}$. Moreover, we incorporated excluded-volume correction arising due to hard-core baryons. We further assumed that the mesons can overlap and fuse into one another and hence cannot generate any hard-core repulsion. This is one major difference between our model and other models [16]. Moreover, we have demonstrated [22] that our calculations give a very good fit to the experimental ratios of multiplicities and other transport coefficients. The grand canonical partition function in our excluded volume model of HG can be written as follows [20-22, 34]:

\begin{equation}
\begin{split}
ln Z_i^{ex} = \frac{g_i}{6 \pi^2 T}\int_{V_i^0}^{V-\sum_{j} N_j V_j^0} dV
\\
\int_0^\infty \frac{k^4 dk}{\sqrt{k^2+m_i^2}} \frac1{[exp\left(\frac{E_i - \mu_i}{T}\right)+1]}
\end{split}
\end{equation}
where $g_i$ is the degeneracy factor of ith species of baryons,$E_{i}$ is the energy of the particle ($E_{i}=\sqrt{k^2+m_i^2}$), $V_i^0$ is the eigenvolume assigned to each baryon of ith species and hence $\sum_{j}N_jV_j^0$ becomes the total occupied volume where $N_{j}$ represent the total number of baryons of jth species calculated in excluded-volume approach.

We can clearly write Eq.(15) as:

\begin{equation}
ln Z_i^{ex} = V(1-\sum_jn_j^{ex}V_j^0)I_{i}\lambda_{i},
\end{equation}
where $I_{i}$ represents the integral:
\begin{equation}
I_i=\frac{g_i}{6\pi^2 T}\int_0^\infty \frac{k^4 dk}{\sqrt{k^2+m_i^2}} \frac1{\left[exp(\frac{E_i}{T})+\lambda_i\right]},
\end{equation}
and $\lambda_i = exp(\frac{\mu_i}{T})$ is the fugacity of the particle, $n_i^{ex}$ is the number density after excluded-volume correction and can be obtained from Eq.(16) as :
\begin{equation}
n_i^{ex} = \frac{\lambda_i}{V}\left(\frac{\partial{ln Z_i^{ex}}}{\partial{\lambda_i}}\right)_{T,V}.
\end{equation}
The total pressure of the HG in our model is [22, 35-36]:
\begin{equation}
\it{p}_{HG}^{ex} = T(1-R)\sum_iI_i\lambda_i + \sum_j\it{p}_j^{meson}.
\end{equation}
In Eq. (19), the second term on the right hand side gives the total summed pressure from all the mesons here taken as pointlike particles. $R=\sum_in_i^{ex}V_i^0$ gives the fractional occupied volume due to all types of baryons. The energy density of HG is obtained from the following relation :
\begin{equation}
\epsilon_{HG}^{ex}=\sum_{i}\left(\frac{T^{2}}{V}\frac{\partial lnZ_i^{ex}}{\partial T}+\mu_{i} n_{i}^{ex}\right)+ \sum_{j}\epsilon_j^{meson}. 
\end{equation}
Similarly entropy density in our model is [22]:
\begin{equation}
s=\frac{\epsilon_{HG}^{ex}+p_{HG}^{ex}-\mu_{B}n_{B}-\mu_{S} n_{S}}{T}.
\end{equation}
Similarly the baryonic succeptibility at $\mu_{B}=0$ can be calculated from the following equation [15]: 
\begin{equation}
\chi^{B}_{2}= T^{2}\left(\frac{\partial^{2} p_{HG}^{ex}/T^{4}}{\partial \mu_{B}^{2}}\right)_{\mu_{B}=\mu_{S}=0}.
\end{equation}
Our calculation for the shear viscosity is completely based on the method outlined by Gorenstein et al. [37]. According to molecular kinetic theory, we can write the dependence of the shear viscosity as follows [37]:
\begin{equation}
\eta \propto n\;l\;\langle|{\bf k}|\rangle , 
\end{equation}
where $n$ is the particle density, $l$ is the mean free path, and hence the average thermal momentum of the baryons or antibaryons is:
\begin{equation}
\langle|{\bf k}|\rangle= \frac{\int_{0}^{\infty}k^{2}\;dk \;k \;{\bf A}}{\int_{0}^{\infty}k^{2}\;dk\;{\bf A}}, 
\end{equation}
and ${\bf A}$ is the Fermi-Dirac distribution function for baryons (anti-baryons). 
For the mixture of particle species with different masses and with the same hard-core radius $r$, the shear viscosity can be calculated by using equation [22, 38]:
\begin{equation}
  \eta=\frac{5}{64 \sqrt{8} \;r^2}\sum_{i}\langle|{\bf k_{i}}|\rangle\times \frac{n_{i}}{n},
\end{equation}
where $n_{i}$ is the number density of the ith species of baryons (or anti-baryons) and $n$ is the total baryon density. 

In order to calculate the speed of sound at constant  $s/n$, we have used the method given by Cleymans and Worku [39]. The speed of sound at $\mu_{B}=0$ is easy to calculate since it is sufficient to keep the temperature as constant [40, 41]. However, the speed of sound $(c_{s})$ at finite chemical potential can be obtained by using the extended expression [39]:

\begin{equation}
c_{s}^{2}=\frac{\left(\frac{\partial p}{\partial T} \right)+ \left(\frac{\partial p} {\partial \mu_{B}} \right)\left(\frac{d\mu_{B}}{dT} \right)+\left(\frac{\partial p} {\partial \mu_{s}} \right)\left(\frac{d\mu_{s}}{dT} \right)}{\left(\frac{\partial \epsilon}{\partial T} \right)+ \left(\frac{\partial \epsilon} {\partial \mu_{B}} \right)\left(\frac{d\mu_{B}}{dT} \right)+\left(\frac{\partial \epsilon} {\partial \mu_{s}} \right)\left(\frac{d\mu_{s}}{dT} \right)},
\end{equation}
where the derivative $d\mu_{B}/dT$ and $d\mu_{s}/dT$ can be evaluated by using two conditions, firstly by keeping $s/n$ constant, and then imposing overall strangeness neutrality [22, 39]:
\begin{equation}
\frac{d\mu_{B}}{dT}=\frac{\left[n\left(\frac{\partial s}{\partial \mu_{s}}\right)-s\left(\frac{\partial n}{\partial \mu_{s}}\right) \right] \left[\frac{\partial L}{\partial T}-\frac{\partial K}{\partial T}\right] -\left[n\left(\frac{\partial s}{\partial T}\right)-s\left(\frac{\partial n}{\partial T}\right)\right]\left[\frac{\partial L}{\partial \mu_{s}}-\frac{\partial K}{\partial \mu_{s}}\right]} {\left[n\left(\frac{\partial s}{\partial \mu_{B}}\right)-s\left(\frac{\partial n}{\partial \mu_{B}}\right)\right]\left[\frac{\partial L}{\partial \mu_{s}}-\frac{\partial K}{\partial \mu_{s}}\right] -\left[n\left(\frac{\partial s}{\partial \mu_{s}}\right)-s\left(\frac{\partial n}{\partial \mu_{s}}\right)\right]\left[\frac{\partial L}{\partial \mu_{B}}-\frac{\partial K}{\partial \mu_{B}}\right]},
\end{equation}
and
\begin{equation}
\frac{d\mu_{s}}{dT}=\frac{\left[n\left(\frac{\partial s}{\partial T}\right)-s\left(\frac{\partial n}{\partial T}\right) \right] \left[\frac{\partial L}{\partial \mu_{B}}-\frac{\partial K}{\partial \mu_{B}}\right] -\left[n\left(\frac{\partial s}{\partial \mu_{B}}\right)-s\left(\frac{\partial n}{\partial \mu_{B}}\right)\right]\left[\frac{\partial L}{\partial T}-\frac{\partial K}{\partial T}\right]} {\left[n\left(\frac{\partial s}{\partial \mu_{B}}\right)-s\left(\frac{\partial n}{\partial \mu_{B}}\right)\right]\left[\frac{\partial L}{\partial \mu_{s}}-\frac{\partial K}{\partial \mu_{s}}\right] -\left[n\left(\frac{\partial s}{\partial \mu_{s}}\right)-s\left(\frac{\partial n}{\partial \mu_{s}}\right)\right]\left[\frac{\partial L}{\partial \mu_{B}}-\frac{\partial K}{\partial \mu_{B}}\right]},
\end{equation}
where $L=n_{s}^{B}+n_{s}^{M}$, is the sum of the strangeness density of baryons and mesons in the HG. Similarly $K=n_{s}^{\bar{B}}+n_{s}^{\bar{M}}$, stands for the sum of anti-strangeness density of baryons and mesons. In all the above calculations, we have taken an equal eigen-volume $V^{0}=\frac{4 \pi r^3}{3}$ for each baryon with a hard-core radius $r=0.8$ fm. Many authors have used hard-core radii varying in the range of $0.5$ fm and $1$ fm [42]. However, Cleymans et al. have used $r = 0.8$ fm in their analysis [43]. It has also been found that proton has approximately exponentially decaying positive charge distribution with an effective mean radius of 0.8 fm [44]. Furthermore, if we consider the energy density existing inside a proton as $4B$ where $B$ is the Bag-constant, then we can write : $4B=M_{p}/V^{0}$ where $M_{p}$ is the mass of a proton. Taking $B^{1/4}=170$ MeV or $B=109 MeV/fm^{3}$, we will get $r\approx 0.8$ fm. We have taken the contributions of all baryons and mesons and their resonances having masses upto $2 GeV/c^{2}$ in our calculation for the HG pressure. We have also used the condition of strangeness neutrality by putting $\sum_{i}S_{i}(n_{i}^{s}-\bar{n}_{i}^{s})=0$, where $S_{i}$ is the strangeness quantum number of the ith hadron, and $n_{i}^{s}(\bar{n}_{i}^{s})$ is the strange (anti-strange) hadron density, respectively [45]. \\

\section{Results and Comparison with Lattice QCD}
\begin{figure}
\includegraphics[height=24em]{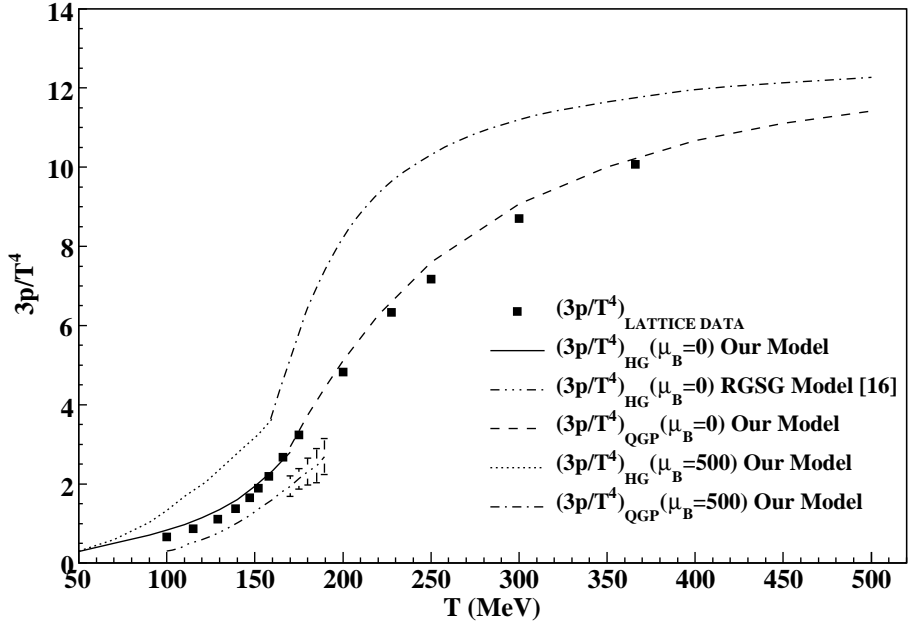}
\caption[]{Variation of 3 times of normalized pressure with respect to temperature at two different values of $\mu_{B}=0$ and $500$ MeV in our hybrid model. Lattice data points at $\mu_{B}=0$ are taken from Ref. [11, 23]. Solid line shows our result in our excluded volume model of HG and dashed curve gives result in quasiparticle model at $\mu_{B}=0$. Dotted curve represents our HG result and dash-dotted curve shows our results in quasiparticle model, at $\mu_{B}=500$ MeV. Dash-tripple dotted curve presents the results obtained by Andronic et. al. by using RGSG model for HG [16].}
\end{figure}

\begin{figure}
\includegraphics[height=24em]{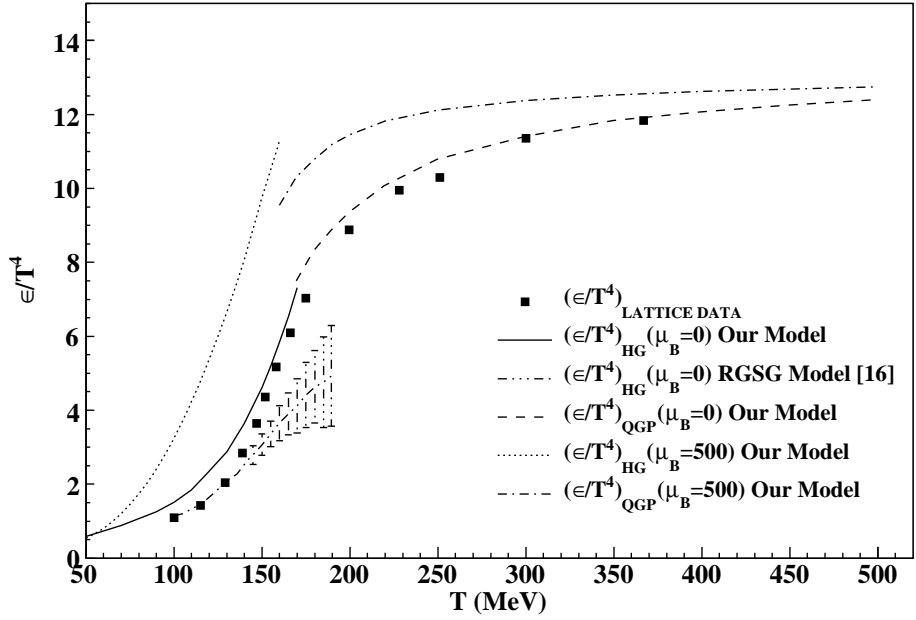}
\caption[]{Variation of normalized energy density with respect to temperature at $\mu_{B}=0$ and $500$ MeV in our hybrid model. Lattice data points at $\mu_{B}=0$ are taken from Refs. [11, 23]. Shaded portion of a curve shows the calculation of Andronic et al [16]. Short-dashed curve with circular marker and long-dashed curve with circular marker presents our HG model results with hard-core radius ($r$) of each baryon as $0.6$ fm at $\mu_{B}=0$ and $500$ MeV, respectively.}
\end{figure}
\begin{figure}
\includegraphics[height=24em]{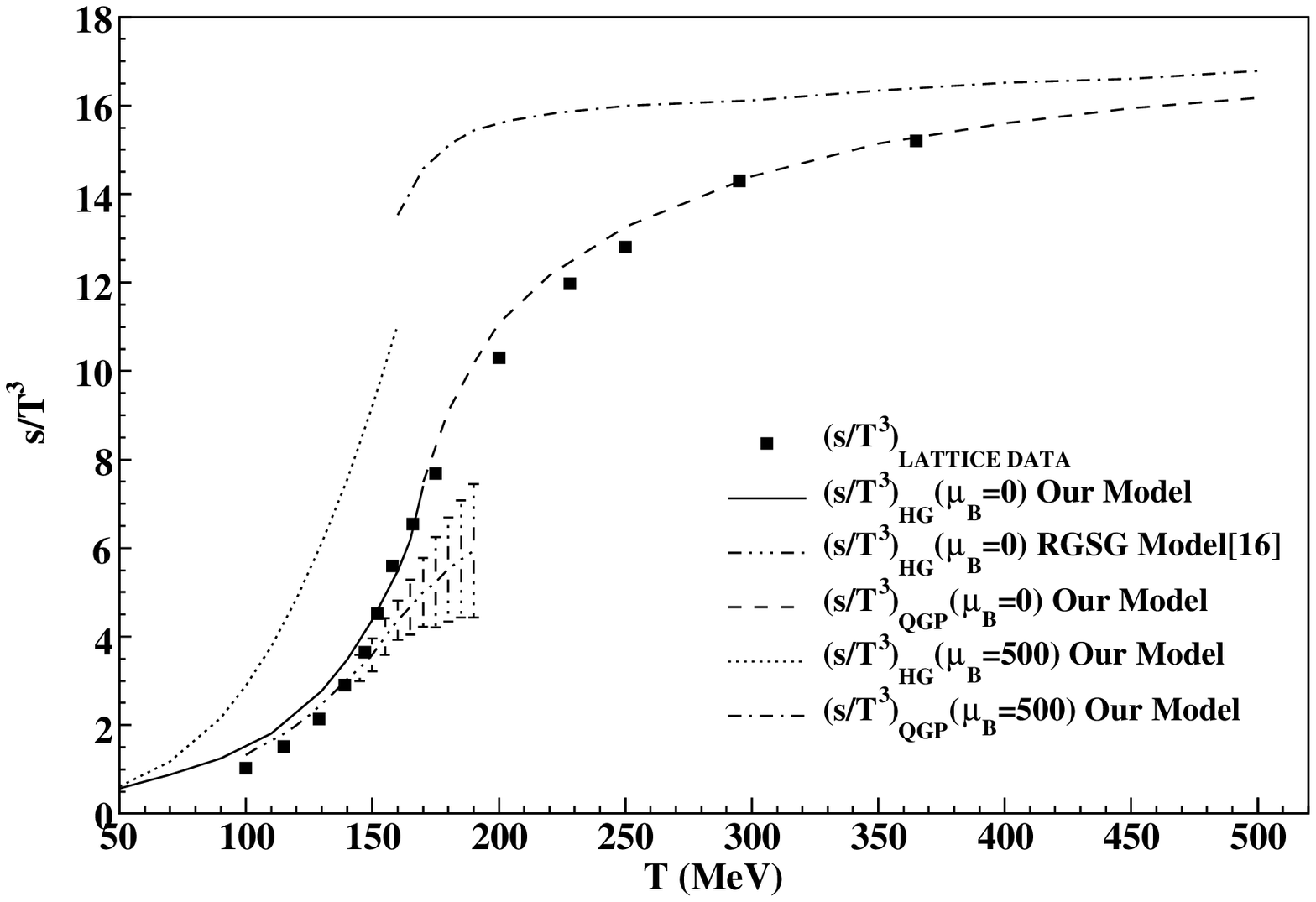}
\caption[]{Variation of normalized entropy density with respect to temperature at $\mu_{B}=0$ and $500$ MeV in our hybrid model. Lattice data points at $\mu_{B}=0$ MeV are taken from Ref. [11, 23]. Shaded portion of a curve represents the calculation of Andronic et al [16].}
\end{figure}

In Fig. 1, we have studied the variations of the quantity $3p/T^{4}$ with temperature at two values of baryon chemical potential $\mu_{B}=0$ and $\mu_{B}=500$ MeV, respectively. At $\mu_{B}=0$, we have also shown the recent lattice results [11, 23] for comparison. We find that our results from the hybrid model are in excellent agreement with the lattice datapoints. For comparison, we have also given the results obtained by Andronic et al. [16] for the low temperature HG phase where they have used the excluded volume model of Rischke, Gorenstein, Stocker and Greiner (RGSG) and they have used a hard-core radius same for each hadron (i.e. meson and baryon) as $r=0.3\pm 0.05$ fm. We notice that our curve shows much better agreement than their curve when compared with the lattice data. We also find that at $\mu_{B}=0$, the curve from our excluded-volume model is smoothly connected with the curve obtained in the quasiparticle calculation around $T=170$ MeV. This result yields enormous faith in the use of the hybrid model and the values of the parameters in our model appear suitably adjusted. Here we stress that we are not using any additional parameters except those which were incorporated in our previous papers [20, 21].

In Fig. 2, we have plotted our results for the variations of normalized energy density ($\epsilon/T^{4}$) with respect to temperature at two values of $\mu_{B}$ i.e., $\mu_{B}=0$ and $\mu_{B}=500$ MeV. For vanishing $\mu_{B}$, our hybrid model results again compare well with the lattice results. Here again the curves obtained in two models of the hybrid model are smoothly connected at $T=170$ MeV but at finite $\mu_{B}$ ($=500$ MeV), we notice a discontinuity in the curves obtained for both the phases of QCD matter. This explains that at $\mu_{B}=0$ MeV and $T=170$ MeV, first-order phase transition does not occur whereas the discontinuity at $\mu_{B}=500$ MeV, indicates the presence of a latent heat in the transition. We have also shown the values of $\epsilon/T^{4}$ at low temperature obtained by Andronic et al. by using RGSG model for HG phase and we observed that our result again shows better agreement with the lattice data in comparison to the results obtained by them. Furthermore, we have also shown the results obtained from excluded volume model for HG if baryons have a hard-core radius $r =0.6$ fm, in order to show the stability of our HG model results. We observe that the change in hard-core radius does not produce any noticeable change in the results.

\begin{figure}
\includegraphics[height=24em]{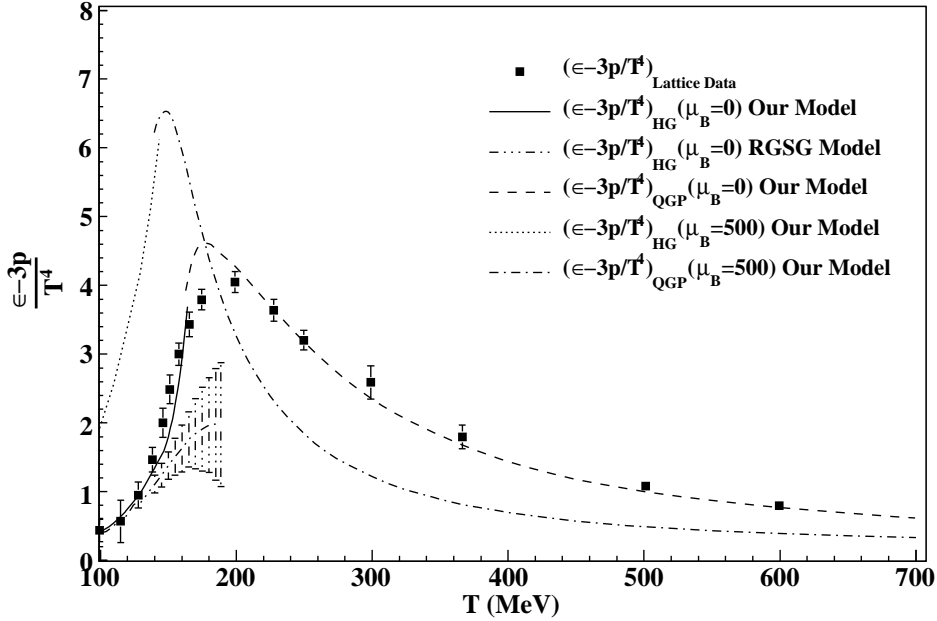}
\caption[]{Variations of trace anomaly $\left(\epsilon-3p\right)/T^{4}$ with respect to temperature at $\mu_{B}=0$ and $500$ MeV in our hybrid model. Lattice points at $\mu_{B}=0$ are taken from Refs. [13, 17, 24]. Shaded portion represents the calculation of Andronic et al. [16].}
\end{figure}
 
In Fig. 3, we show the variations of normalized entropy density ($s/T^{3}$) with temperature at two values of $\mu_{B}$ (i.e., $0$ and $500$ MeV), respectively. For $\mu_{B}=0$ MeV, our hybrid model yields good agreement with the lattice results. Here again the quantity $s/T^{3}$ is smoothly connected at vanishing $\mu_{B}$ when we view the results of HG and QGP models but a disconnected graph appears at $\mu_{B}=500$ MeV. Dashed-tripple dotted curve enclosing a shaded portion shows the results obtained by Andronic et. al [16] using an excluded volume model for the HG where the hard-core radius for each hadron (meson and baryon) is taken as $r=0.3\pm 0.05$ fm and this gives rise to a shaded portion. At finite $\mu_{B}$ ($=500$ MeV), the discontinuity in $s/T^{3}$ curves for both the phases shows the presence of latent heat involved in the phase transition from HG to QGP.

In Fig. 4, we have plotted the results obtained for the trace anomaly factor $\left(\epsilon-3p\right)/T^{4}$ in our hybrid model calculations using HG and QGP equations of state separately at $\mu_{B}=0$. We further compare our results with the results obtained in a recent lattice calculation [13, 17, 24]. We notice that our results yield an excellent fit to the lattice data. The success of our hybrid model which involves a separate and distinct description for both the phases (i.e., low temperature HG and large temperature QGP), is indeed excellent in reproducing the features of the lattice curves. We also show the results at finite baryon chemical potential i.e. $\mu_{B}=500$ MeV and it shows that the peak of trace anomaly factor is found to shift towards the lower temperature side. One other important thing we observe is that by introducing the finite baryon chemical potential into the system the curves show a completely different behaviour at low as well as high temperature side. We also present here the result calculated by Andronic et. al. by dashed-triple dotted shaded curve [16] and we notice that our results yield better fit to the data.

\begin{figure}
\includegraphics[height=24em]{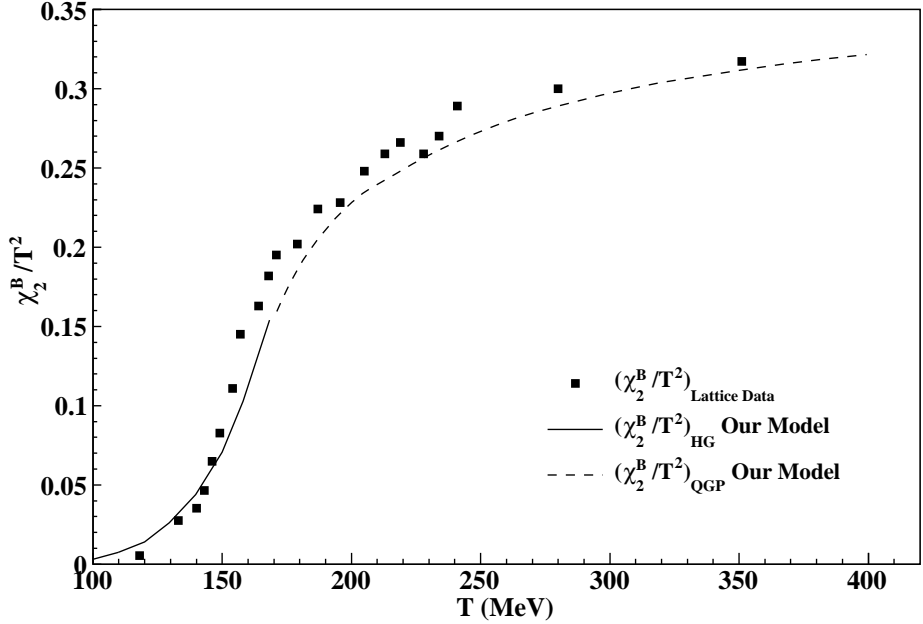}
\caption[]{Variation of normalized baryon number succeptibility with respect to temperature at $\mu_{B}=0$ in our hybrid model. Lattice data points at $\mu_{B}=0$ are taken from Ref. [24].}
\end{figure}

\begin{figure}
\includegraphics[height=24em]{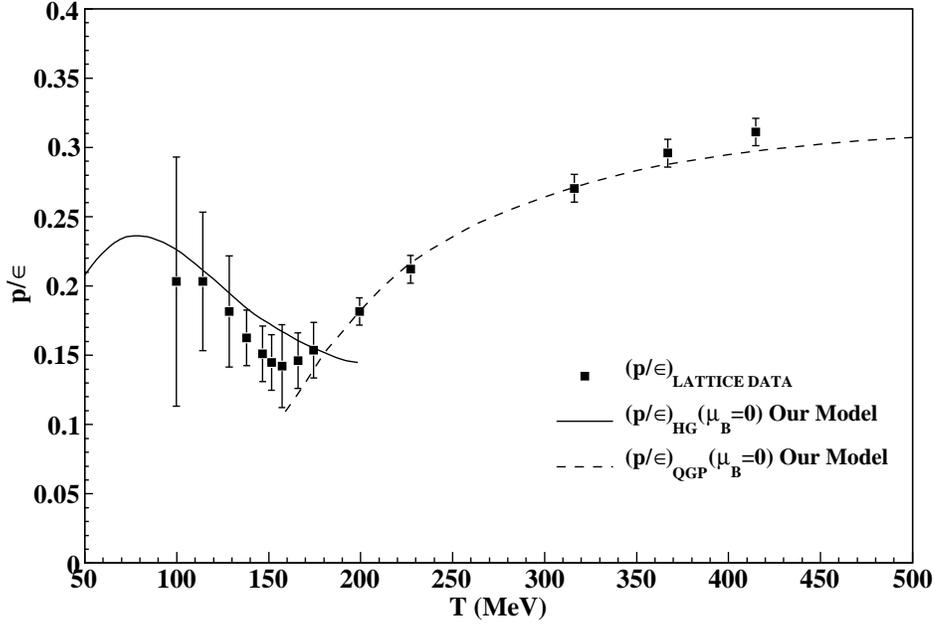}
\caption[]{Variation of $p/\epsilon$ with respect to temperature at $\mu_{B}=0$ in our hybrid model. Lattice data points at $\mu_{B}=0$ are taken from Refs. [11, 23].}
\end{figure}

\begin{figure}
\includegraphics[height=24em]{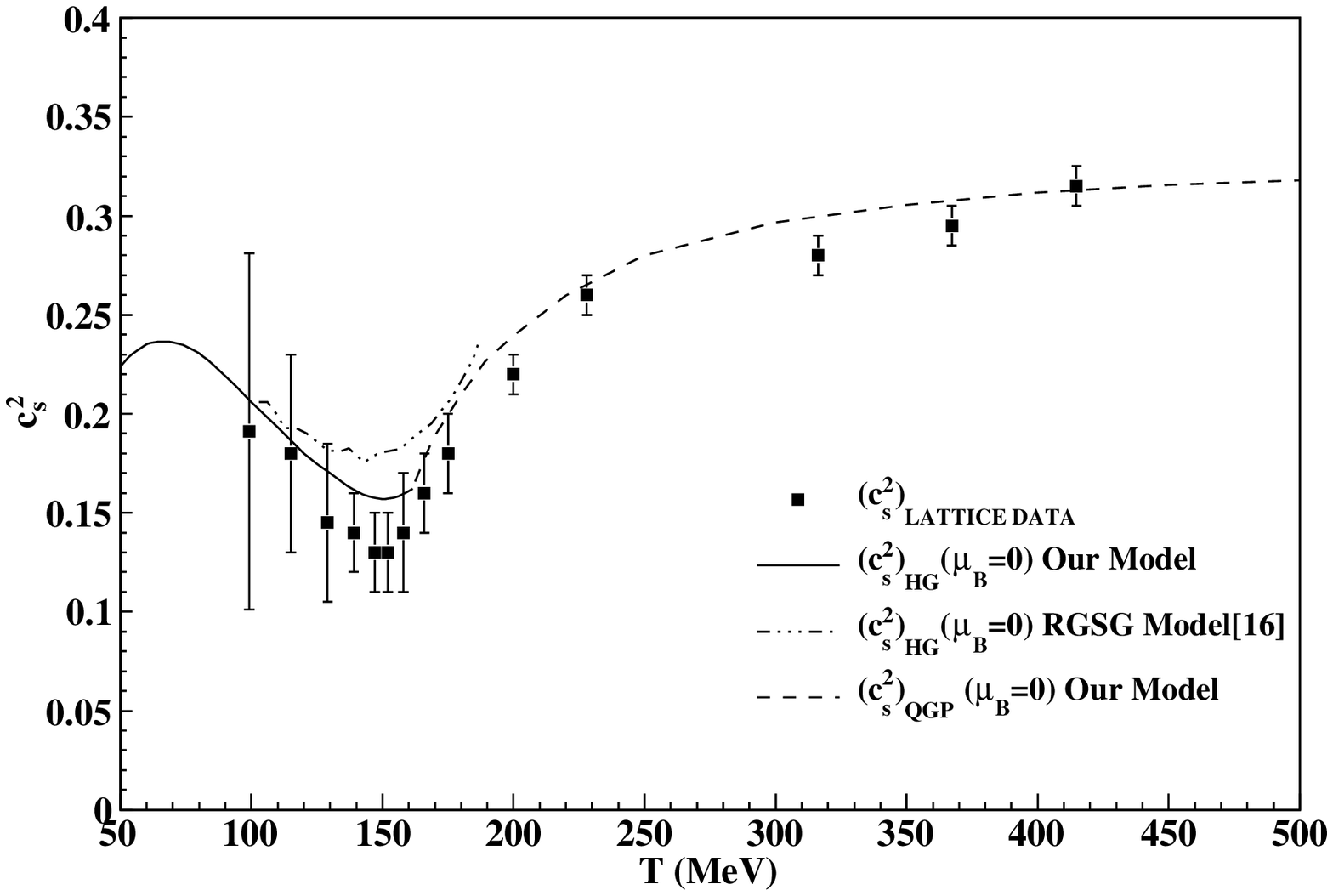}
\caption[]{Variation of speed of sound with the temperature at $\mu_{B}=0$ MeV as obtained in our hybrid model and compared with lattice results [11, 23]. Dash-tripple dotted curve shows the result of Andronic et al. [16].}
\end{figure}

In Fig. 5, we show the variations of baryonic succeptibility normalized as $\chi^{B}_{2}/T^{2}$ with temperature at $\mu_{B}=0$ and compare our results with the lattice QCD results. For $\mu_{B}=0$ our hybrid model results again compare well with the lattice data points. Here again the curves, for $\chi^{B}_{2}/T^{2}$ obtained for HG and QGP phases are smoothly connected at around $T=170$ MeV, which shows the presence of a cross-over transition between two phases. In Fig. 6, we show the variations of pressure to energy density ratio with respect to temperature at $\mu_{B}=0$ and we find that our results are well supported by the lattice QCD results. We can view a discontinuity in the behaviour of this curve when we go from low temperature phase to high temperature phase and the discontinuity and/or a minimum appears at around $T=170$ MeV.  

In Fig. 7, we show the variations of square of speed of sound ($c_{s}^{2}$) with respect to temperature at $\mu_{B}=0$ MeV and again a comparison is given with the lattice QCD results. For $\mu_{B}=0$, our results reproduces the lattice results very well. Here again the quantity $c_{s}^{2}$ shows a similar discontinuity and/or minimum where both the curves for both phases i.e., HG and QGP join each other and, therefore it further supports a smooth crossover transition between two phases. We have also shown separately the curve for $c_{s}^{2}$ obtained by Andronic et. al. by using RGSG model for HG phase only [16]. However, the features of the curve differ from the lattice data, although it also yields a minimum at around the same temperature.

\section{Critical Point (CP) and Order of Phase Transition}

 In this section we attempt to show the use of hybrid model in constructing a deconfining phase boundary between HG and QGP by using Gibbs' criteria and fixing the precise location and nature of CP existing on this phase boundary.

\begin{figure}
\includegraphics[height=28em]{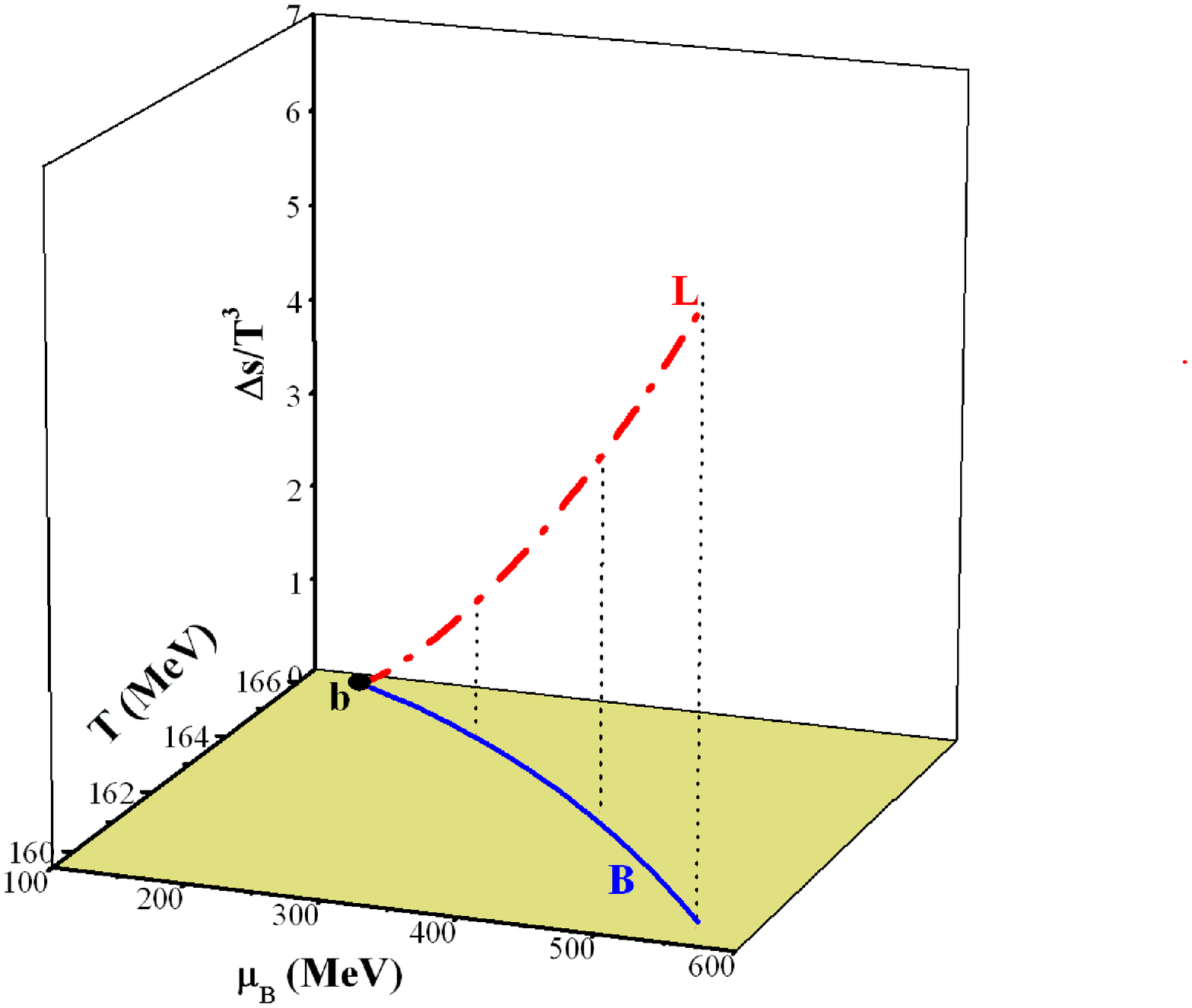}
\caption[]{Variation of $ (\Delta s/T^{3})=(s/T^{3})_{QGP}- (s/T^{3})_{HG}$ with respect to coordinates of various phase transition points on the $(T, \mu_{B})$ phase boundary. We have used transition points from Ref. [20].}
\end{figure}

\begin{figure}
\includegraphics[height=28em]{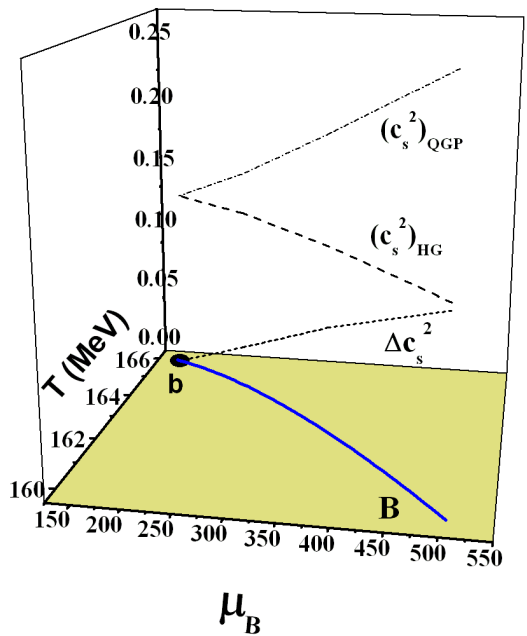}
\caption[]{Variation of $c_{s}^{2})_{QGP}$ (dash-dotted line ), $(c_{s}^{2})_{HG}$ (dashed line), and $ (\Delta c_{s}^{2})=(c_{s}^{2})_{QGP}- (c_{s}^{2})_{HG}$ (short-dashed line) with respect to coordinates of various phase transition points on the $(T, \mu_{B})$ phase boundary. We have used transition points from Ref. [20].}
\end{figure}

\begin{figure}
\begin{center}
\includegraphics[height=24em]{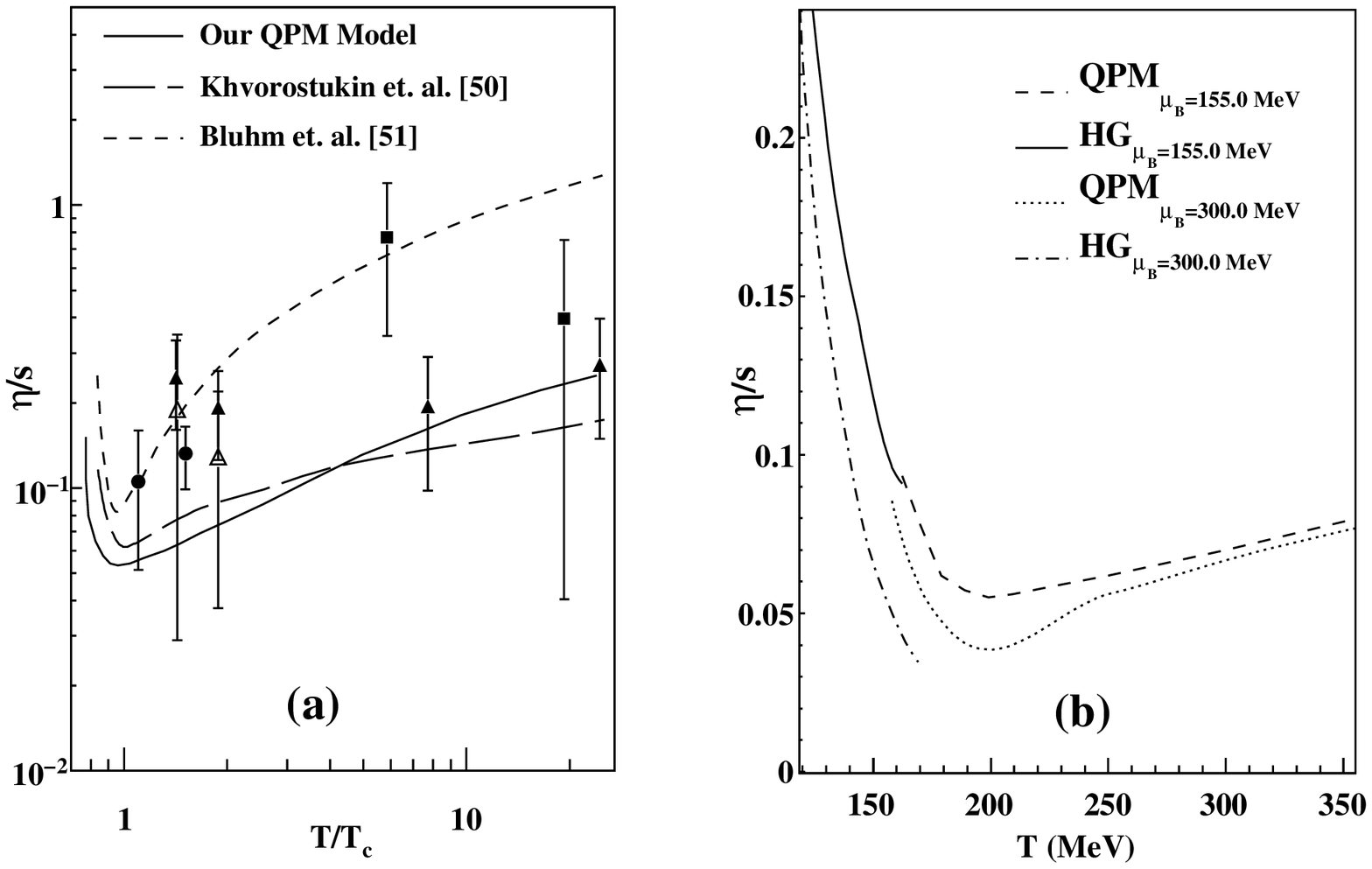}
\caption[]{(a) Variations of shear viscosity to entropy density ratio ($\eta/s$) for gluon plasma with respect to $T/T_{c}$. Solid line is the result of our calculation. Long-dashed line is the result obtained in Ref. [53] and short-dashed line is taken from Ref. [54]. The lattice data with $16^{3}\times 8$ and $24^{3}\times 8$ lattice are from Refs. [55] (triangles and squares) and [56] (filled circles). (b) Variations of $\eta/s$ obtained in our calculation for HG and QGP phases with temperature at different values of $\mu_{B}$.}
\end{center}
\end{figure}

In Fig. 8, we show what will happen to a quantity depicting the change in the entropy density from the phase transition at CP of the phase diagram. We define the normalized difference $\frac{\Delta s}{T^{3}}=(s/T^{3})_{QGP}-(s/T^{3})_{HG}$ and demonstrate its variations with respect to the coordinates of the phase transition points lying at the deconfining phase boundary. We find that $\frac{\Delta s}{T^{3}}\ne 0.0$ and positive along the deconfining phase boundary in the case of first order phase transition which supports the absorption of a nonvanishing latent heat in the phase transition from HG to QGP. However, we surprisingly notice that $\frac{\Delta s}{T^{3}}\approx 0$ exactly at the CP and thus CP can be taken as a point where the first order phase boundary terminates and phase transition changes its order. 

Sound velocity yields an important transport property of the QCD matter created in the nucleus-nucleus collision experiments because the hydrodynamic evolution of this matter strongly depends on it. Speed of sound is related to the speed of small perturbations produced in the QCD matter in its local rest frame. A minimum in $c_{s}$ has also been interpreted in terms of a phase transition point [39-40, 46-51] where a large number of degrees of freedom present in the medium causes difficulty in its propagation. Further, Chojnacki and Florkowski [52] have proposed that a shallow minimum in the speed of sound near the smooth joining of HG and QGP phases in a hybrid model description corresponds to the presence of a cross-over transition. They have obtained a temperature dependence in the sound velocity by exploiting QCD lattice simulation for the high temperature phase (i.e. $T > 1.5 T_{c}$) and ideal hadron gas description at low temperature (i.e. $T < 0.15 T_{c}$) and used further interpolations to connect them smoothly. In Fig. 9, we have separately shown the variation of square of speed of sound i.e., $c_{s}^{2}$ for HG and QGP media. We have also shown the difference $(\Delta c_{s}^{2})=(c_{s}^{2})_{QGP}- (c_{s}^{2})_{HG}$ and demonstrated its variations with respect to the coordinates of the phase transition points defining the deconfining phase boundary. We again find that $\Delta c_{s}^{2}\approx 0.0$ at CP. Thus this result lends further support to our finding regarding the change of the order of the transition at the critical point.

In Fig. 10 (a), we plot the variation of $\eta/s$ of gluon plasma with respect to $T/T_{c}$, where $T_{c}$ is the value of the critical temperature. We compare our model result with the results obtained by Khvorostukin et. al. [53] and Bluhm et. al. [54]. A comparison with lattice calculations [55-56] is also demonstrated in Fig. 10 (a). We observe that the results obtained in our calculation agrees well with the lattice data even at very large temperatures. However, the error quoted in the lattice simulation is quite large. Our result also appears in close agreement with the result obtained by Khvorostukin et. al [53] while the results of Bluhm et al [54] neither agrees with the lattice data nor shows any agreement with our calculations. It should be mentioned here that the authors [53-54] have used the QPM to calculate shear viscosity of a gluon plasma. The difference appears when they use the relaxation time ($\tau$) and its dependence on the strong coupling constant ($g$). Khvorostukin et al [53] have taken $\tau\propto g^{2}$ while Bluhm et al [54] have assumed $\tau\propto G^{4}$ where $G$ is an effective strong coupling constant. This result naturally gives us additional confidence in the use of QPM for an EOS of QGP. In Fig. 10 (b), we plot the variations of shear viscosity to entropy density ratio for HG and QGP, separately with temperature at $\mu_{B}=300$ MeV (dashed-line) and  at the critical potential $\mu_{c}=155$ MeV (solid-line). At $\mu_{B}=300$ MeV, we observe a discontinuity in $\eta/s$ at the joining point of the curves for both the phases. Further we observe an upward jump in $\eta/s$ as we go from low temperature HG phase to high temperature QGP phase and this supports the result obtained by Sasaki and Redlich [26]. At critical chemical potential  $\mu_{c}=155$ MeV and temperature $T_{c}=166$ MeV [20-21], we get a cusp like behaviour in $\eta/s$ while going from HG to QGP phase as pointed out in Ref. [26, 57].
 
 Thus above results give a firm indication that the order of phase transition changes at CP for the deconfinement phase transition. It should be added here that many authors in the past have also used two separate equations of state for QGP and HG and obtained a tentative explanation for an analytic and smooth cross-over and CP in their models [58-60]. Our model presents a similar picture. Here we explicitly and separately consider both the phases, i.e., HG as well as QGP and hence it gives a clear understanding how a first-order deconfining phase transition can be constructed in nature. At almost vanishing baryon density, overlapping mesons fuse into each other and form a large bag or cluster, whereas at high baryon density, hard-core repulsion among baryons restricts the mobility of baryons in the HG. Consequently we consider two distinct limiting regimes of HG, one is a meson-dominant while the other is a baryon dominant region and CP occurs exactly at the joining of two regimes in our models.

Searching for the precise location and the nature of the critical point (CP) on the QCD phase diagram are still a challenging problem. Although various calculations have predicted its existence but the quantitative predictions regarding its location wildly differ. Experiments face an uphill task in probing the CP in QCD phase diagram because a clarity in theoretical prediction is missing. Moreover, many unstudied problems such as short lifetime and the reduced volume of the QGP formed at colliders also affect the location of CP and its verification [61]. In these circumstances, we hope that our results will clarify the mist surrounding the understanding of the deconfining phase transition. More importantly, we have formulated a phenomenological hybrid model which provides a realistic EOS for the entire QCD matter and in the absence of the first-principle lattice QCD calculation especially at finite $\mu_{B}$, it can be reliably used for deriving the information on the QCD phase boundary.\\

\noindent
{\bf Acknowledgments}\\

 PKS is grateful to the University Grants Commission (UGC), New Delhi for financial assistance.

\pagebreak


\begin{thebibliography}{99}

\bibitem{[1]} C. P. Singh, Phys. Rep. 236, 147 (1993); Int. J. Mod. Phys. A7, 7185 (1992). 
\bibitem{[2]} A. Andronic et. al. Nucl. Phys. A837, 65-86 (2010).
\bibitem{[3]} J. Cleymans, R. V. Gavai, E. Suhonen, Phys. Rep.130, 217 (1986).
\bibitem{[4]} M. A. Stephanov, Int. J. Mod. Phys. A20, 4387 (2005); Prog. Theor. Phys. Suppl. 153, 139 (2004).
\bibitem{[5]} M. A. Stephanov, Phys. Rev. Lett. 102, 032301 (2009).
\bibitem{[6]} M. A. Stephanov, K. Rajagopal and E. V. Shuryak, Phys. Rev. Lett. 81, 4816 (1998).
\bibitem{[7]} R. V. Gavai and S. Gupta, Phys. Rev. D71, 114014 (2005).
\bibitem{[8]} Z. Fodor and S. D. Katz, J. High Energy Phys. 0404, 050 (2004); Y. Aoki et. al., Nature 443, 675 (2006).
\bibitem{[9]} P. de Forcrand and O. Philipsen, J. High Energy Phys. 01, 077 (2007); J. High Energy Phys. 11, 012 (2008).
\bibitem{[10]} Owe Philipsen, arXiv:1111.5370v1 [hep-ph].
\bibitem{[11]} P. Huovinen, P. Petreczky, Nucl. Phys. A837, 26 (2010); Journal of Phys.: Conference Series 230, 012012 (2010).
\bibitem{[12]} P. Huovinen, P. Petreczky, C. Schmidt, arXiv:1202.3104v1 [nucl-th].
\bibitem{[13]} C. Ratti et. al., Nucl. Phys. A855, 253 (2011).
\bibitem{[14]} A. Tawfik, Phys. Rev. D71, 054502 (2005).
\bibitem{[15]} A. Bazavov et. al., HotQCD Collaboration, arXiv:1203.0784v1[hep-lat].
\bibitem{[16]} A. Andronic, P. Braun-Munzinger, J. Stachel, M. Winn, arXiv:1201.0693v1 [nucl-th]; D. H. Rischke, M. I. Gorenstein, H. Stocker and W. Greiner, Z. Phys. C51, 485 (1991).
\bibitem{[17]} S. Plumari, W. M. Alberico, V. Greco, and C. Ratti, Phys. Rev. D84, 094004 (2011).
\bibitem{[18]} M. Bluhm, B. Kampfer, and G. Soff, Phys. Lett. B620, 131 (2005); M. Bluhm, B. Kampfer, R. Schulze, D. Seipt, and U. Heinz, Phys. Rev. C76, 034901 (2007).
\bibitem{[19]} M. Bluhm and B. Kampfer, Phys. Rev. D77, 034004 (2008); 77,114016 (2008); W. Cassing, Nucl. Phys. A795, 70 (2007).
\bibitem{[20]} P. K. Srivastava, S. K. Tiwari and C. P. Singh, Phys. Rev. D82, 014023 (2010).
\bibitem{[21]} P. K. Srivastava, S. K. Tiwari and C. P. Singh, Nucl. Phys. A 862-863CF, 424 (2011).
\bibitem{[22]} S. K. Tiwari, P. K. Srivastava, and C. P. Singh, Phys. Rev. C85, 014908 (2012).
\bibitem{[23]} S. Borsanyi et. al., JHEP 1011, 077 (2010).
\bibitem{[24]} S. Borsanyi et. al. J. Phys. G: Nucl. Part. Phys. 38, 124060 (2011); JHEP 1201, 138 (2012)
\bibitem{[25]} L. P. Csernai, J. I. Kapusta and L. D. McLerran, Phys. Rev. Lett. 97, 152303 (2006).
\bibitem{[26]} C. Sasaki, K. Redlich, Nucl. Phys. A 832, 62 (2010).
\bibitem{[27]} R. A. Lacey et. al., Phys. Rev. Lett. 98, 092301 (2007).
\bibitem{[28]} V. M. Bannur, Phys. Lett. B647, 271 (2007); J. Phys. G: Nucl. Part. Phys. 32, 993 (2006); Eur. Phys. J. C50, 629-634 (2007); Phys Rev. C78, 045206 (2008).
\bibitem{[29]} M. I. Gorenstein and S. N. Yang, Phys. Rev. D52, 5206 (1995).
\bibitem{[30]} Min He, J.-F. Li, W.-M.Sun, and H.-S. Zong, Phys. Rev. D79, 036001 (2009).
\bibitem{[31]} C. Sasaki and K. Redlich, Phys. Rev. C 79, 055207 (2009).
\bibitem{[32]} A. Hosoya and K. Kajantie, Nucl. Phys. B250, 666 (1985).
\bibitem{[33]} P. Arnold, G. D. Moore, and L. G. Yaffe, J. High Energy Phys. 11, 001 (2000); J. High Energy Phys. 05, 051 (2003)
\bibitem{[34]} C. P. Singh, P. K. Srivastava and S. K. Tiwari, Phys. Rev. D80, 114508 (2009); Erratum-ibid. D 83: 039904 (2011).
\bibitem{[35]} S. Uddin and   C. P. Singh, Zeit. f. Phys. C63, 147 (1994).
\bibitem{[36]} C. P. Singh, B. K. Patra and K. K. Singh, Phys. Lett. B387, 680 (1996).
\bibitem{[37]} M. I. Gorenstein, M. Hauer, and O. N. Moroz, Phys. Rev. C 77, 024911 (2008).
\bibitem{[38]} E. M. Lifschitz and L. P. Pitaevski, Physical Kinetics, 2nd ed. (Pergamon Press, Oxford, 1981), Chap.1, p.3.
\bibitem{[39]} J. Cleymans and D. Worku, Mod. Phys. Lett. A 26, 1197 (2011).
\bibitem{[40]} P. Castorina, J. Cleymans, D. E. Miller, H. Satz, Eur. Phys. J. C 66, 207 (2010).
\bibitem{[41]} P. Braun-Munzinger, J. Stachel, Nucl. Phys. A606, 320 (1996).
\bibitem{[42]} L. J. Reinders, H. Rubinstein, S. Yasaki, Phys. Rep. 127, 1 (1985); S. O. B$\ddot{a}$ckman, G. E. Brown, J. A. Niskanen, Phys. Rep. 124, 1 (1985).
\bibitem{[43]} J. Cleymans et. al., Z. Phys. C33, 151 (1986).
\bibitem{[44]} J. -L. Basdevant, J. Rich, M. Spiro, Fundamentals in Nuclear Physics, (Springer, 2005), p. 155.
\bibitem{[45]} M. Mishra and C. P. Singh, Phys. Rev. C78, 024910 (2008); Phys. Lett. B651, 119 (2007).
\bibitem{[46]} R. V. Gavai, A. Gocksch, Phys. Rev. D 33, 614 (1986).
\bibitem{[47]} K. Redlich, H. Satz, Phys. Rev. D 33, 3747 (1986).
\bibitem{[48]} F. Karsch, PoS CPOD07, 026 (2007).
\bibitem{[49]} P. Braun-Munzinger, J. Stachel, Nucl. Phys. A 606, 320 (1996).
\bibitem{[50]} D. Prorok, L. Turko, arXiv:hep-ph/0101220.
\bibitem{[51]} J. Noronha-Hostler, J. Noronha, and C. Greiner, Phys. Rev. Lett. 103, 172302 (2009).
\bibitem{[52]} M. Chojnacki and W. Florkowski, Acta Phys. Polo. B 38, 3249 (2007).
\bibitem{[53]} A. S. Khvorostukhin, V. D. Toneev, and D. N. Voskresensky, Phys. Rev. C 83, 035204 (2011).
\bibitem{[54]} M. Bluhm, B. Kampfer, and K. Redlich, Nucl. Phys. A 830, 737c (2009).
\bibitem{[55]} S. Sakai and A. Nakamura, PoS LAT 2077, 221 (2007).
\bibitem{[56]} H. B. Meyer, Phys. Rev. D 76, 101701(R) (2007).
\bibitem{[57]} P. Castorina, K. Redlich, and H. Satz, Eur. Phys. J. C 59, 67 (2009).
\bibitem{[58]} A. S. Kapoyannis, Eur. Phys. J. C51, 135 (2007); N. G. Antoniou, A. S. Kapoyannis,Phys. Lett. B563, 165 (2003).
\bibitem{[59]} K. A. Bugaev, Phys. Rev. C76, 014903 (2007).
\bibitem{[60]} O. Lourenco, M. Dutra, A. Delfino, and M. Malheiro, Phys. Rev. D84, 125034 (2011).
\bibitem{[61]} A. Gopie and M. C. Ogilvie, Phys. Rev. D59, 034009 (1999); O. Kiriyama and A. Hosaka, Phys. Rev. D67, 085010 (2003); L. F. Palhares, E. S. Fraga, and T. Kodama, J. Phys. G 38, 085101 (2011).




 \end{thebibliography}
\end{document}